\documentclass[aps,pre]{revtex4}

\begin{document}

\title{Open Questions on Nonequilibrium Thermodynamics of Chemical Reaction Networks}

\author{Massimiliano Esposito}
\affiliation{Complex Systems and Statistical Mechanics, Department of Physics and Materials Science, University of Luxembourg, L-1511 Luxembourg, Luxembourg}

\date{\today}

\begin{abstract}
Chemical reaction networks (CRNs) are prototypical complex systems because reactions are nonlinear and connected in intricate ways, and they are also essential to understand living systems. 
Here, I discuss how recent developments in nonequilibrium thermodynamics provide new insight on how CRNs process energy and perform sophisticated tasks, and describe open challenges in the field.

\end{abstract}
\maketitle


Thermodynamics is a powerful tool to study complex systems. 
It predicts for instance that the performance of any thermal machines, regardless of its details, cannot surpass Carnot efficiency. 
But traditional thermodynamics is an equilibrium theory that makes no reference to time and thus to dynamics.
Phenomenological attempts to incorporate time started with linear irreversible thermodynamics \cite{Prigogine1961Jul} and later lead to finite-time-thermodynamics \cite{Bejan1996Feb}, respectively over the first and second half of the 20th century.
But it is only since the 21st century that kinetics and energetics have been systematically merged together into a consistent theory called stochastic thermodynamics (ST) \cite{Seifert2012}.

Interestingly, chemical systems always played a key role in the development of thermodynamics (e.g. in the work of J. W. Gibbs, T. de Donder, L. Onsager, I. Progogine) and ST in not an exception. 
Elements of the theory were already discovered during the second half of the 20th century in the context of the stochastic description of chemical reactions \cite{Hill1977,Schnakenberg1976,Nicolis1984} and, as we will see, thermodynamics of chemical reaction networks (CRNs) is still driving new developments. 

In its modern formulation \cite{Rao2018Feb}, ST constructs thermodynamic observables (e.g. heat, work, dissipation) on top of the stochastic dynamics of an open system. The key assumption is that all the degrees of freedom which are not explicitly described in the dynamics (i.e. the internal structure of the system states) must be at equilibrium. The link between observables and dynamics is then provided by the local detailed balance property: when an elementary process (e.g. an elementary reaction) induces a transition between two states (e.g. the reactants and products), the Boltzmann constant $k_B$ times the log ratio of the forward and backward current between the two states is the total dissipation of the process, i.e. the entropy dissipated into the environment plus the entropy change in the system. 
The resulting theory forms a superstructure built on top of the dynamics which can be used to formulate no-go theorems, e.g. no dynamics can implement operations forbidden by the second law. 

Furthermore, by analytically or numerically solving a specific dynamics, ST predicts the temporal changes of the thermodynamic quantities and can be used to address typical finite-time thermodynamics questions such as identifying driving protocols leading to maximum power extraction.   
This has been applied for instance to molecular motors, Brownian ratchets and electronic circuits (for reviews see e.g. \cite{Ciliberto2017Jun, Benenti2017Jun}).

ST also predicts the nonequilibrium fluctuations of thermodynamic observables. 
The central finding of the field was the discovery of fluctuation relations asserting that the probability to observe a given positive total dissipation is exponentially more likely then the probability of its negative counterpart. 
This result generalizes the second law which only states that on average total dissipation cannot decrease. 
Previous results such as Onsager reciprocity relations or the fluctuation-dissipation theorem are recovered from it when considering close to equilibrium situations. 

ST also provides very natural connections with information theory. 
The nonequilibrium system entropy contains a Shannon entropy contribution, dissipation can be expressed as a Kullback-Leibler entropy quantifying how different probabilities of forward trajectories are from their time-reversed counterpart, and the difference between the nonequilibrium and equilibrium thermodynamic potential take the form of Kullback-Leibler entropies between the system nonequilibrium and equilibrium probability distribution. 
These results provided a rigorous ground to assess the cost of various information processing operations such as Landauer erasure or Maxwell demons \cite{Parrondo2015Feb}.

More recently, ST has also been used to show that dissipation can be used to set universal bounds (called thermodynamic uncertainty relations) on the precision \cite{Horowitz2020} as well as on the duration \cite{falasco2020dissipationtime} of a process. 

When turning to CRNs, a key ingredient that makes ST of CRN particularly rich is the existence of a nontrivial macroscopic limit giving rise to nonlinear dynamics caused by entropic effects whenever bimolecular reactions are involved.
But before proceeding, a number of points should be clarified.

First, a priori only elementary reactions should be considered to ensure the validity of the local detailed balance property of the rates. 

Second, the distinction between closed and open CRNs is essential. 
The former correspond to elementary reactions in a closed vessel that must and will eventually reach equilibrium.
The latter instead correspond to an open vessel in which the concentrations or the in- and out-flow of some species are externally controlled.
These chemostatted species will usually prevent the system from reaching equilibrium by continuously performing chemical work on the system.

Third, CRNs can be described using a stochastic or a deterministic dynamics. 
The former describes the CRN in terms of probabilities of all molecular counts evolving in time according to a chemical master equation.
The latter correspond to the macroscopic limit of the former and describes the CRN in terms of concentrations obeying rate equations satisfying mass action law.

Forth, linear and nonlinear CRNs are very different.
Linear CRNs are made of effectively unimolecular reactions, where ``effectively'' is used to emphasize that a linear reaction can be obtained by externally fixing (i.e. chemostatting) the concentration of one of the two species involved in a bimolecular reaction.  
These CRNs have the peculiarity that the stochastic and deterministic description coincide: the concentration of a species can be interpreted as a probability or as an average molecular count. In other words the macroscopic limit of linear CRNs is trivial and can be pictured as Gaussian fluctuations centering around their average. 
This is not the case for nonlinear CRNs which can give rise to complex dynamical behaviors (multistabilities, oscillations and chaos) absent for linear CRNs.
Here the deterministic solutions can be seen as mean field solutions which captures the most likely behaviors of the stochastic description in the limit of large molecular numbers, but the fluctuations can take very complex forms, for instance multimodal in case of multistability.  

By making use of the concepts developed in ST, a nonequilibrium thermodynamics of CRNs encompassing the aforementioned points was recently developed for both deterministic and stochastic dynamics \cite{Rao2016, Rao2018b}. A crucial outcome is that the topology of the CRN (defined by the stoichiometric matrix specifying how reactions are coupled to each other, and by the identity of the chemostatted species) determines 1) the number and specific form of the thermodynamic forces (i.e. combinations of chemical potentials of the chemostatted species) which drive the molecular currents across the CRN and contribute to the dissipation and 2) the thermodynamic potential quantifying the free energy stored inside the CRN.  
This newly developed theory opens many interesting research avenues to be explored in coming years.  

{\it Energy transduction} and its performance was extensively studied in simple linear CRNs which are well suited to describe the various conformation changes of systems such as enzymes or molecular motors \cite{Seifert2012,Ge2012Jan}. 
But the newly developed tools opened the way to performance studies of energy storage and transduction in simple synthetic nonlinear CRNs \cite{Penocchio2019a}. 
The next step is to move towards a systematic formulation of energy transduction in complex nonlinear networks such as metabolic networks. 
Analyzing the interplay between topology and thermodynamics could provide novel insight into the organization and performance of metabolism. 

{\it Chemical fluctuations} have been studied in nonlinear CRNs displaying nonequilibrium phase transitions, but mostly using minimal models chosen to push analytical calculation as far as possible \cite{Gaspard2004,Lazarescu2019}.
The methods used (which are already nontrivial) need to be extended to study realistic biochemical networks performing interesting tasks. 
Biochemical complexity emerges at the subcellular level in part because some species with very low molecular number coexists with others with many orders of magnitude more molecules. 
The former provide the versatility (via fluctuations) while the latter the stability (via deterministic attractors) necessary for a cell to reliably switch between different modes of operation.           
To be able to analyze such mechanisms and assess their reliability and energetic cost, a thermodynamics for CRNs described by mixed dynamics (i.e. dynamics which can treat some species stochastically and other deterministically) needs to be developed.

{\it Complex chemical signals} play a key role to transmit or process information in biology and we are now in a position to assess the thermodynamic cost needed to generate them.
Recent studies on complex spatio-temporal chemical signals such as Turing patterns or chemical waves evaluated the chemical work that needs to be supplied by the chemostats to produce structures (quantified in terms of information theoretic metrics) that would otherwise be washed off by diffusion \cite{Falasco2018a, Avanzini2019a}.
But here too, minimal models handpicked for tractability were used and more realistic setups must now be considered.

{\it Computing} in the traditional sense of carrying out logic circuits or machine learning can be done with artificial CRNs. 
At the same time, many operations implemented by biological CRNs (e.g. sensing, copying and proofreading) can also be considered as forms of computation \cite{Grozinger2019Nov}. 
Thermodynamics of CRNs provides powerful methods to explore the trade-offs that may exist between cost, accuracy, precision and speed of chemical computing operations. 
It could also help assess the fundamental limits of chemical computation and possibly formulate meaningful comparisons with electronic computation.

{\it Thermodynamically consistent coarse graining} is a crucial challenge for thermodynamics of CRNs. 
Indeed, the need to resolve elementary reactions is a priori a severe practical limitation for biochemical applications. 
Fortunately, important results have been obtained (at steady state) showing for instance that ST can be straightforwardly applied to describe non-elementary reactions catalyzed by cytosolic enzymes \cite{Wachtel2018}. 
This is very good news for studies on metabolism, but more work is needed to consider transient effects and find out if ST could also be used to assess the cost of more complex networks such as genetic ones.

I have only sketched some of the promising research directions enabled by the recent developments in nonequilibrium thermodynamics of CRN, and many more could have been mentioned (e.g. nonisothermal effects, non-ideal solutions, photo- and electro-chemistry).
But I hope that I managed to convey the message that thermodynamics of CRNs is a very promising field that will become indispensable to the study of chemical complexity.
I expect that much progress will happen in the years to come which has the potential to impact our understanding of metabolism, chemical information processing, origins of life, and biogeochemistry.

\section*{Competing interests}
The author declares no competing interests.

\section*{Acknowledgments}
The author is funded by the European Research Council project NanoThermo (ERC-2015-CoG Agreement No. 681456).


\begin{thebibliography}{22}
\expandafter\ifx\csname natexlab\endcsname\relax\def\natexlab#1{#1}\fi
\expandafter\ifx\csname bibnamefont\endcsname\relax
  \def\bibnamefont#1{#1}\fi
\expandafter\ifx\csname bibfnamefont\endcsname\relax
  \def\bibfnamefont#1{#1}\fi
\expandafter\ifx\csname citenamefont\endcsname\relax
  \def\citenamefont#1{#1}\fi
\expandafter\ifx\csname url\endcsname\relax
  \def\url#1{\texttt{#1}}\fi
\expandafter\ifx\csname urlprefix\endcsname\relax\def\urlprefix{URL }\fi
\providecommand{\bibinfo}[2]{#2}
\providecommand{\eprint}[2][]{\url{#2}}

\bibitem[{\citenamefont{Prigogine}(1961)}]{Prigogine1961Jul}
\bibinfo{author}{\bibfnamefont{I.}~\bibnamefont{Prigogine}},
  \emph{\bibinfo{title}{Introduction to Thermodynamics of Irreversible
  Processes, 2nd Ed.}} (\bibinfo{publisher}{Interscience Publishers, New York},
  \bibinfo{year}{1961}).

\bibitem[{\citenamefont{Bejan}(1996)}]{Bejan1996Feb}
\bibinfo{author}{\bibfnamefont{A.}~\bibnamefont{Bejan}}, \bibinfo{journal}{J.
  Appl. Phys.} \textbf{\bibinfo{volume}{79}}, \bibinfo{pages}{1191}
  (\bibinfo{year}{1996}), ISSN \bibinfo{issn}{0021-8979}.

\bibitem[{\citenamefont{Seifert}(2012)}]{Seifert2012}
\bibinfo{author}{\bibfnamefont{U.}~\bibnamefont{Seifert}},
  \bibinfo{journal}{Rep. Prog. Phys.} \textbf{\bibinfo{volume}{75}},
  \bibinfo{pages}{126001} (\bibinfo{year}{2012}),
  \urlprefix\url{https://doi.org/10.1088/0034-4885/75/12/126001}.

\bibitem[{\citenamefont{Hill}(1977)}]{Hill1977}
\bibinfo{author}{\bibfnamefont{T.}~\bibnamefont{Hill}},
  \emph{\bibinfo{title}{Free Energy Transduction in Biology}}
  (\bibinfo{publisher}{Academic Press}, \bibinfo{year}{1977}).

\bibitem[{\citenamefont{Schnakenberg}(1976)}]{Schnakenberg1976}
\bibinfo{author}{\bibfnamefont{J.}~\bibnamefont{Schnakenberg}},
  \bibinfo{journal}{Rev. Mod. Phys.} \textbf{\bibinfo{volume}{48}},
  \bibinfo{pages}{571} (\bibinfo{year}{1976}),
  \urlprefix\url{https://link.aps.org/doi/10.1103/RevModPhys.48.571}.

\bibitem[{\citenamefont{Jiu-li et~al.}(1984)\citenamefont{Jiu-li, Van~den
  Broeck, and Nicolis}}]{Nicolis1984}
\bibinfo{author}{\bibfnamefont{L.}~\bibnamefont{Jiu-li}},
  \bibinfo{author}{\bibfnamefont{C.}~\bibnamefont{Van~den Broeck}},
  \bibnamefont{and} \bibinfo{author}{\bibfnamefont{G.}~\bibnamefont{Nicolis}},
  \bibinfo{journal}{Z. Physik B - Condensed Matter}
  \textbf{\bibinfo{volume}{56}}, \bibinfo{pages}{165} (\bibinfo{year}{1984}),
  \urlprefix\url{https://doi.org/10.1007/BF01469698}.

\bibitem[{\citenamefont{Rao and Esposito}(2018{\natexlab{a}})}]{Rao2018Feb}
\bibinfo{author}{\bibfnamefont{R.}~\bibnamefont{Rao}} \bibnamefont{and}
  \bibinfo{author}{\bibfnamefont{M.}~\bibnamefont{Esposito}},
  \bibinfo{journal}{New J. Phys.} \textbf{\bibinfo{volume}{20}},
  \bibinfo{pages}{023007} (\bibinfo{year}{2018}{\natexlab{a}}), ISSN
  \bibinfo{issn}{1367-2630}.

\bibitem[{\citenamefont{Ciliberto}(2017)}]{Ciliberto2017Jun}
\bibinfo{author}{\bibfnamefont{S.}~\bibnamefont{Ciliberto}},
  \bibinfo{journal}{Phys. Rev. X} \textbf{\bibinfo{volume}{7}},
  \bibinfo{pages}{021051} (\bibinfo{year}{2017}), ISSN
  \bibinfo{issn}{2160-3308}.

\bibitem[{\citenamefont{Benenti et~al.}(2017)\citenamefont{Benenti, Casati,
  Saito, and Whitney}}]{Benenti2017Jun}
\bibinfo{author}{\bibfnamefont{G.}~\bibnamefont{Benenti}},
  \bibinfo{author}{\bibfnamefont{G.}~\bibnamefont{Casati}},
  \bibinfo{author}{\bibfnamefont{K.}~\bibnamefont{Saito}}, \bibnamefont{and}
  \bibinfo{author}{\bibfnamefont{R.~S.} \bibnamefont{Whitney}},
  \bibinfo{journal}{Phys. Rep.} \textbf{\bibinfo{volume}{694}},
  \bibinfo{pages}{1} (\bibinfo{year}{2017}), ISSN \bibinfo{issn}{0370-1573}.

\bibitem[{\citenamefont{Parrondo et~al.}(2015)\citenamefont{Parrondo, Horowitz,
  and Sagawa}}]{Parrondo2015Feb}
\bibinfo{author}{\bibfnamefont{J.~M.~R.} \bibnamefont{Parrondo}},
  \bibinfo{author}{\bibfnamefont{J.~M.} \bibnamefont{Horowitz}},
  \bibnamefont{and} \bibinfo{author}{\bibfnamefont{T.}~\bibnamefont{Sagawa}},
  \bibinfo{journal}{Nat. Phys.} \textbf{\bibinfo{volume}{11}},
  \bibinfo{pages}{131} (\bibinfo{year}{2015}), ISSN \bibinfo{issn}{1745-2481}.

\bibitem[{\citenamefont{Horowitz and Gingrich}(2020)}]{Horowitz2020}
\bibinfo{author}{\bibfnamefont{J.~M.} \bibnamefont{Horowitz}} \bibnamefont{and}
  \bibinfo{author}{\bibfnamefont{T.~R.} \bibnamefont{Gingrich}},
  \bibinfo{journal}{Nat. Phys.} \textbf{\bibinfo{volume}{16}},
  \bibinfo{pages}{15} (\bibinfo{year}{2020}),
  \urlprefix\url{https://doi.org/10.1038/s41567-019-0702-6}.

\bibitem[{\citenamefont{Falasco and
  Esposito}(2020)}]{falasco2020dissipationtime}
\bibinfo{author}{\bibfnamefont{G.}~\bibnamefont{Falasco}} \bibnamefont{and}
  \bibinfo{author}{\bibfnamefont{M.}~\bibnamefont{Esposito}},
  \emph{\bibinfo{title}{The dissipation-time uncertainty relation}}
  (\bibinfo{year}{2020}),
  \bibinfo{note}{\href{https://arxiv.org/abs/2002.03234}{arXiv:2002.03234}},
  \urlprefix\url{https://arxiv.org/abs/2002.03234}.

\bibitem[{\citenamefont{Rao and Esposito}(2016)}]{Rao2016}
\bibinfo{author}{\bibfnamefont{R.}~\bibnamefont{Rao}} \bibnamefont{and}
  \bibinfo{author}{\bibfnamefont{M.}~\bibnamefont{Esposito}},
  \bibinfo{journal}{Phys. Rev. X} \textbf{\bibinfo{volume}{6}},
  \bibinfo{pages}{041064} (\bibinfo{year}{2016}),
  \urlprefix\url{https://link.aps.org/doi/10.1103/PhysRevX.6.041064}.

\bibitem[{\citenamefont{Rao and Esposito}(2018{\natexlab{b}})}]{Rao2018b}
\bibinfo{author}{\bibfnamefont{R.}~\bibnamefont{Rao}} \bibnamefont{and}
  \bibinfo{author}{\bibfnamefont{M.}~\bibnamefont{Esposito}},
  \bibinfo{journal}{J. Chem. Phys.} \textbf{\bibinfo{volume}{149}},
  \bibinfo{pages}{245101} (\bibinfo{year}{2018}{\natexlab{b}}),
  \urlprefix\url{https://doi.org/10.1063/1.5042253}.

\bibitem[{\citenamefont{Ge et~al.}(2012)\citenamefont{Ge, Qian, and
  Qian}}]{Ge2012Jan}
\bibinfo{author}{\bibfnamefont{H.}~\bibnamefont{Ge}},
  \bibinfo{author}{\bibfnamefont{M.}~\bibnamefont{Qian}}, \bibnamefont{and}
  \bibinfo{author}{\bibfnamefont{H.}~\bibnamefont{Qian}},
  \bibinfo{journal}{Phys. Rep.} \textbf{\bibinfo{volume}{510}},
  \bibinfo{pages}{87} (\bibinfo{year}{2012}), ISSN \bibinfo{issn}{0370-1573}.

\bibitem[{\citenamefont{Penocchio et~al.}(2019)\citenamefont{Penocchio, Rao,
  and Esposito}}]{Penocchio2019a}
\bibinfo{author}{\bibfnamefont{E.}~\bibnamefont{Penocchio}},
  \bibinfo{author}{\bibfnamefont{R.}~\bibnamefont{Rao}}, \bibnamefont{and}
  \bibinfo{author}{\bibfnamefont{M.}~\bibnamefont{Esposito}},
  \bibinfo{journal}{Nat. Commun.} \textbf{\bibinfo{volume}{10}},
  \bibinfo{pages}{3865} (\bibinfo{year}{2019}),
  \urlprefix\url{https://doi.org/10.1038/s41467-019-11676-x}.

\bibitem[{\citenamefont{Gaspard}(2004)}]{Gaspard2004}
\bibinfo{author}{\bibfnamefont{P.}~\bibnamefont{Gaspard}}, \bibinfo{journal}{J.
  Chem. Phys.} \textbf{\bibinfo{volume}{120}}, \bibinfo{pages}{8898}
  (\bibinfo{year}{2004}), \eprint{https://doi.org/10.1063/1.1688758},
  \urlprefix\url{https://doi.org/10.1063/1.1688758}.

\bibitem[{\citenamefont{Lazarescu et~al.}(2019)\citenamefont{Lazarescu,
  Cossetto, Falasco, and Esposito}}]{Lazarescu2019}
\bibinfo{author}{\bibfnamefont{A.}~\bibnamefont{Lazarescu}},
  \bibinfo{author}{\bibfnamefont{T.}~\bibnamefont{Cossetto}},
  \bibinfo{author}{\bibfnamefont{G.}~\bibnamefont{Falasco}}, \bibnamefont{and}
  \bibinfo{author}{\bibfnamefont{M.}~\bibnamefont{Esposito}},
  \bibinfo{journal}{J. Chem. Phys.} \textbf{\bibinfo{volume}{151}},
  \bibinfo{pages}{064117} (\bibinfo{year}{2019}),
  \eprint{https://doi.org/10.1063/1.5111110},
  \urlprefix\url{https://doi.org/10.1063/1.5111110}.

\bibitem[{\citenamefont{Falasco et~al.}(2018)\citenamefont{Falasco, Rao, and
  Esposito}}]{Falasco2018a}
\bibinfo{author}{\bibfnamefont{G.}~\bibnamefont{Falasco}},
  \bibinfo{author}{\bibfnamefont{R.}~\bibnamefont{Rao}}, \bibnamefont{and}
  \bibinfo{author}{\bibfnamefont{M.}~\bibnamefont{Esposito}},
  \bibinfo{journal}{Phys. Rev. Lett.} \textbf{\bibinfo{volume}{121}},
  \bibinfo{pages}{108301} (\bibinfo{year}{2018}),
  \urlprefix\url{https://link.aps.org/doi/10.1103/PhysRevLett.121.108301}.

\bibitem[{\citenamefont{Avanzini et~al.}(2019)\citenamefont{Avanzini, Falasco,
  and Esposito}}]{Avanzini2019a}
\bibinfo{author}{\bibfnamefont{F.}~\bibnamefont{Avanzini}},
  \bibinfo{author}{\bibfnamefont{G.}~\bibnamefont{Falasco}}, \bibnamefont{and}
  \bibinfo{author}{\bibfnamefont{M.}~\bibnamefont{Esposito}},
  \bibinfo{journal}{J. Chem. Phys.} \textbf{\bibinfo{volume}{151}},
  \bibinfo{pages}{234103} (\bibinfo{year}{2019}),
  \urlprefix\url{https://doi.org/10.1063/1.5126528}.

\bibitem[{\citenamefont{Grozinger et~al.}(2019)\citenamefont{Grozinger, Amos,
  Gorochowski, Carbonell, Oyarz{\ifmmode\acute{u}\else\'{u}\fi}n, Stoof,
  Fellermann, Zuliani, Tas, and
  Go{\ifmmode\tilde{n}\else\~{n}\fi}i-Moreno}}]{Grozinger2019Nov}
\bibinfo{author}{\bibfnamefont{L.}~\bibnamefont{Grozinger}},
  \bibinfo{author}{\bibfnamefont{M.}~\bibnamefont{Amos}},
  \bibinfo{author}{\bibfnamefont{T.~E.} \bibnamefont{Gorochowski}},
  \bibinfo{author}{\bibfnamefont{P.}~\bibnamefont{Carbonell}},
  \bibinfo{author}{\bibfnamefont{D.~A.}
  \bibnamefont{Oyarz{\ifmmode\acute{u}\else\'{u}\fi}n}},
  \bibinfo{author}{\bibfnamefont{R.}~\bibnamefont{Stoof}},
  \bibinfo{author}{\bibfnamefont{H.}~\bibnamefont{Fellermann}},
  \bibinfo{author}{\bibfnamefont{P.}~\bibnamefont{Zuliani}},
  \bibinfo{author}{\bibfnamefont{H.}~\bibnamefont{Tas}}, \bibnamefont{and}
  \bibinfo{author}{\bibfnamefont{A.}~\bibnamefont{Go{\ifmmode\tilde{n}\else\~{n}\fi}i-Moreno}},
  \bibinfo{journal}{Nat. Commun.} \textbf{\bibinfo{volume}{10}},
  \bibinfo{pages}{1} (\bibinfo{year}{2019}), ISSN \bibinfo{issn}{2041-1723}.

\bibitem[{\citenamefont{Wachtel et~al.}(2018)\citenamefont{Wachtel, Rao, and
  Esposito}}]{Wachtel2018}
\bibinfo{author}{\bibfnamefont{A.}~\bibnamefont{Wachtel}},
  \bibinfo{author}{\bibfnamefont{R.}~\bibnamefont{Rao}}, \bibnamefont{and}
  \bibinfo{author}{\bibfnamefont{M.}~\bibnamefont{Esposito}},
  \bibinfo{journal}{New J. Phys.} \textbf{\bibinfo{volume}{20}},
  \bibinfo{pages}{042002} (\bibinfo{year}{2018}),
  \urlprefix\url{https://doi.org/10.1088/1367-2630/aab5c9}.

\end{thebibliography}
\end{document}